\documentclass[prl,aps,twocolumn,superscriptaddress,showpacs]{revtex4}

\usepackage{amsbsy}
\usepackage{amssymb}
\usepackage[dvips]{graphicx}

\begin{document}

\newcommand \be  {\begin{equation}}
\newcommand \bea {\begin{eqnarray} \nonumber }
\newcommand \ee  {\end{equation}}
\newcommand \eea {\end{eqnarray}}

\title{{\bf Finite-size effects and intermittency in a simple aging system}}

\author{Estelle Pitard}
\affiliation{Laboratoire des Verres (CNRS-UMR n$^o$5587), CC69, Universit\'{e}
Montpellier 2, 34095 Montpellier Cedex 5, France.}

\date{\today}

\begin{abstract}

We study the intermittent dynamics and the fluctuations of the dynamic correlation function
of a simple aging system.
Given its size $L$ and its coherence length $\xi$,
the system can be  divided into
$N$ independent subsystems, where $N=(\frac{L}{\xi})^d$,
 and $d$ is the dimension of space.
 Each of them is considered as an aging
subsystem  which evolves according 
to an activated dynamics between energy levels.
 We compute analytically
the distribution of trapping times for the global system, which can take power-law, stretched-exponential or
exponential forms according to the values of $N$ and the regime of times considered. An effective number of
subsystems at age $t_w$, $N_{eff}(t_w)$, can be defined,
which decreases as $t_w$ increases, as well as an effective coherence length, $\xi(t_w) \sim
t_w^{(1-\mu)/d}$, where $\mu <1$ characterizes the trapping times distribution of a single
subsystem.
We also compute the probability distribution functions of the time intervals between large
decorrelations, which exhibit different power-law behaviours as $t_w$ increases (or $N$ decreases), and
which should be accessible experimentally.
 Finally, we calculate the probability distribution function of the two-time correlator.
 We
show that in a phenomenological approach, where $N$ is replaced by
the effective number of subsystems $N_{eff}(t_w)$, the same qualitative behaviour
as in experiments and simulations of several glassy systems can be
obtained.

\end{abstract}


\maketitle

\vskip 1cm

\section{I. Introduction}
\label{sec: intro}

The dynamics of glassy materials such as spin-glasses, structural glasses or amorphous
soft materials like gels, pastes or foams has been a subject of considerable study
\cite{YOUNG, EDIGER, SOFT}. Considerable effort has been made in order to understand and
quantify the out-of-equilibrium character of their temporal relaxation, in particular the
absence of
time translational invariance (aging), through the study of dynamic correlation functions.
On the theoretical side, global dynamic correlations have been described
at a mean-field level in disordered systems such as spin-glasses \cite{REVUE_LETIC}, or at a phenomenological
level in models such as the Random Energy Model \cite{REM,MB}. 
Both approaches neglect all spatial properties of the system, and are 
therefore likely to miss any spatial correlations
that arise during the dynamics. Alternatively, a phenomenological
picture, the droplet model, has been proposed, that focuses on the spatial properties 
as a key to understand the slow dynamics and the critical properties of spin-glasses
\cite{BRAYMOORE, FISHER}.

In recent years, interest in the spatial properties of glassy systems has been growing.
The size of excitations in finite-dimensional spin-glasses has been studied numerically \cite{KRZA}. In
simulations of kinetically constrained glassy systems \cite{RITSOL, KCM},
and of supercooled liquids \cite{DFGP, LACEVI},
cooperativity lengths have been identified, and related to the presence of heterogeneities
in the dynamics
\cite{HARROWELL, GARRAHAN}.

 More generally, a lot of physical
questions remain; a crucial one being: in what manner, at
a microscopic level, does a glassy system evolve, both spatially and temporally?
What are the spatial configurations of the typical rearrangements experienced by a glassy system during its
relaxation and how are they affected by aging?
 Are
there any common relaxation mechanisms of glassy systems, though there also exist
specificities to given materials?
Recently, new results have been obtained in this direction. Focusing on the 
spatial aspect of glassy relaxation, cooperative rearrangements events have been
evidenced, both experimentally \cite{VANBLAAD, WEEKS}
and through computer simulations \cite{KOB}, stimulating new
research on the challenging question of coherence lengths and cooperativity in glasses.
On the  temporal side, beautiful experiments have shown evidence of temporal
intermittency in colloidal gels and micellar polycrystals \cite{CIP},
 and in polycarbonate glasses \cite{CILIBERTO}. 
It seems now well
established that in glasses and gels, relaxation takes place in a discontinuous way, involving
sudden rearrangements followed by periods of arrest where almost nothing happens.
The precise experimental determination of the distribution of time lags between rearranging
events will give insight into the characterictic ``trapping times'' of the system. 
In
experiments on glasses, this distribution seems to be close to a power-law \cite{CILIBERTO} -which is consistent with a
trap model with an exponential distribution of energies \cite{MB}-,
 whereas
it has been found in simulations of 
supercooled glasses to correspond to a model of traps with either an exponentiel
distribution of energies, or a gaussian distribution of energies \cite{DOLIWA,DENNY}.
Therefore, quantities of interest  are not just average quantities,
but also fluctuations, and in particular the full probability 
distributions of correlations. The study of fluctuations in
glassy systems may contain subtle information, as was already realized by Israeloff and Weissman 
\cite{WEISSMAN}, who analyzed carefully
mesoscopic noise in spin-glasses in an attempt to discriminate between a model of droplets, and a scenario of
hierarchical dynamics.

Recently, probability distributions of two-time correlation
functions in gels and glasses
have been shown to exhibit a non-Gaussian behaviour \cite{CIP, CILIBERTO}. These non-Gaussian features have been
found also in numerical simulations of disordered systems and kinetically constrained models
\cite{CHAMON, SIBANI, RITORT_FT}. They have been tentatively 
explained using the  analogy between glassy dynamics and
critical dynamics \cite{CHAMON}, for which universal, non-Gaussian features can be expected \cite{BRAMWELL}.

In this study, we will not advocate any similarity with critical dynamics,
but we will rather try to deduce the non-Gaussian behaviour merely from finite-size effects in a simple
out-of-equilibrium glassy model.

We will consider a system that can be divided into $N$ independent subsystems.
Each subsystem is supposed to represent an independent model of
glassy relaxation between energy traps. Such a model of traps has been studied
extensively \cite{MB, THESEBERTIN}. We will show that
the superposition of the $N$ subsystems will have the same average
dynamical correlation as one individual subsystem; however, its probability distribution
function will depend strongly
on $N$; in the limit $N \rightarrow \infty$, one has to recover the Gaussian distribution, according to the
central limit theorem.
Moreover, the distribution of time intervals between relaxation events or between decorrelations will depend
crucially on the number of subsystems, i.e on the value of the internal coherence length.

The paper is organized as follows. In Section II, we introduce the model and recall the main results of
\cite{MB}. In Section III, we calculate the distribution of time intervals between successive events
in the whole system. In Section IV, we calculate the distribution
of time intervals between successive decorrelations in the system, the same way they can be measured in Time
Resolved Correlation experiments on soft glassy materials \cite{CIP}.
 Section V is devoted to the probability distribution of the two-time correlation
function. Finally, in Section VI, we summarize
and discuss our results.

\section{II. Definition of the model}

Let us consider a  simple model of dynamics of a system
between energy levels. In a generic disordered system, it is reasonable to assume that
low-lying energy levels are exponentially distributed: this is the case for example
for the lowest energy levels (non-extensive corrections to the ground state energy) of the
Random Energy model (REM) \cite{REM,KH} or in spin-glasses \cite{MPV}.
Experimental determination of energy barriers in low-temperature glasses also seem to support this exponential
distribution \cite{LEPORINI}.

More precisely, let us call $E$ an energy barrier, which is the difference between a reference energy level
which is taken as the origin for the energies, and a negative energy level. $E$ is hence a
positive quantity. We choose the
distribution of barriers as
$\rho(E)=\frac{1}{E_0} e^{-E/E_0}$.

Changes in configurations in a disordered or glassy system are often attributed to thermally
activated  events over energy barriers \cite{DOLIWA, DENNY}, although other mecanisms exist,
such as kinetically constrained models \cite{RITSOL}, which do not require an energy landscape,
 and are able
to reproduce some of the features of glassy materials.
In the language of activated events, a trapping time $\tau$ corresponding to a barrier $E$ can
be defined as $\tau=\tau_0 e^{E/k_B T}$, where $\tau_0$ is a microscopic time scale, $k_B$ the
Boltzmann constant, and $T$ the temperature. For an exponential distribution of barriers, the
distribution of trapping times is equal to
$\psi(\tau)=\mu \frac{\tau_0^{\mu}}{\tau^{1+\mu}}$, where $\mu=\frac{k_B T}{E_0}$.

The dynamics of a system evolving in such an energy landscape has been studied by different
authors, according to which choice of transition rates between energy levels is made
\cite{dDO,KH,MB,MAAS}.
In the following, we will consider only the case where the transition rate from barrier $E$ to
any other barrier$E'$ is $W(E \rightarrow E')=\frac{1}{\tau_0} e^{-E/k_B T}$, \cite{KH, MB},
which means that the escape from the initial trap is the limiting process, whatever the
destination.
Note that this family of models does not include any kind of spatial structure, since energy is
not related here to spatial configurations, and that it is mean-field in nature since 
transitions to all levels are allowed with the same probability.
 Extensions of these models to
finite dimensionalities have however been attempted in \cite{BERTIN, MONTHUS}.
We will see in the following that
the  superposition of several  of such systems can actually introduce (though
rather artificially) a relevant lengthscale.

Dynamical properties of the model have been studied in detail in \cite{MB}. In
particular, when $\mu < 1$, the model exhibits aging (absence of time translation invariance of
the correlations), whereas it is time-translationally invariant for $\mu >1$. In this paper, we
will focus on the case $\mu <1$.
For a given trajectory of the system, the two-time correlation function $C(t_w, t_w+\tau)$ is
defined as:
$C(t_w, t_w+\tau)=1$ if the system has remained in the same energy trap between $t_w$ and
$t_w+\tau$, 
$C(t_w, t_w+\tau)=0$ if between $t_w$ and
$t_w+\tau$, the system has left the energy trap it was in at $t_w$.

Averaging over all barrier configurations (which we denote by $<..>$), one obtains the average
two-time correlation function:

$$\Pi(t_w,t_w+\tau) =<C(t_w, t_w+\tau)>$$

It was shown in \cite{MB} that for large $t_w$, this function is given by the following formula:

$$\Pi(t_w,t_w+\tau) \simeq  \frac{\sin{\pi \mu}}{\pi} 
\int_{\frac{\tau}{t_w}(1+\frac{\tau}{t_w})} ^1  dv (1-v)^{\mu -1} v^{-\mu}$$

which we will use in sections III and V.

\subsection{The sprinkling density $S(t)$}

Another important quantity for our study is the sprinkling density of events
at time $t$,  $S(t)$. An ``event'' has to be understood as a jump from an energy
level to another one. The sprinkling density of events  $S(t)$
is defined as the time density distribution of having an event at time $t$ -whatever the number
of events before $t$- and given that there was one event at $t=0$.
This is a standard quantity defined in the context of renewal theory  \cite{FELLER}.
For any distribution of trapping times $\psi(\tau)$ 
where the trapping times are independent
random variables -and given that it does not depend
on the age, i.e on the choice of the time origin-, and independently of the type of dynamics
used, the following formula holds:

\begin{equation}
S(t)=\psi(t)+\int_0^{t} dt_l
 S(t_l) \psi(t-t_l) \label{sprinkle}
\end{equation}
 
\noindent
where $t_l$ stands for the time of the last event to have taken place before
$t$; the first term $\psi(t)$ corresponds to the special case $t_l=0$.

In the following, we will use the following trapping times distribution:
$\psi(\tau)=\frac{\mu \tau_0^{\mu}}{(\tau_0+\tau)^{1+\mu}}$,
in order to ensure that $\tau$ can take values from $0$ to $\infty$.
We will also need the large time behaviour of the corresponding sprinkling density $S(t)$.
Following the lines of \cite{BARDOU}, this can be easily computed using Laplace transforms.
Using the notation ${\hat f}(z)$ to denote the Laplace transform of a function $f(t)$, equation
(1) is equivalent to

$${\hat S}(z)=\frac{{\hat \psi}(z)}{1-{\hat \psi}(z)}$$

The Laplace transform ${\hat \psi}(z)$ can be computed as:

$$ {\hat \psi}(z)=\int_0^{\infty} dt e^{-zt} \psi(t)
=\mu \tau_0^{\mu} e^{ \tau_0 z} z^{\mu} \int_{\tau_0  z}^{\infty} dv \frac{e^{-v}}{v^{-1-\mu}}$$

Two cases have to be considered before taking the limit $\tau_0 z \rightarrow 0$.

 If $\mu < 1$,
${\hat \psi}(z) = 1-\Gamma(1-\mu) (\tau_0 z)^{\mu} +
\frac{1}{1-\mu} \tau_0 z + o(\tau_0 z)$.

\noindent
Then ${\hat S}(z) \simeq \frac{1}{\Gamma(1-\mu)}\frac{1}{(\tau_0 z)^{\mu}}$,
and for $t \gg \tau_0$, 
$S(t) \simeq  c(\mu) \frac{t^{\mu -1}}{\tau_0^{\mu}}$; with
$c(\mu)=\frac{1}{\Gamma(1-\mu) \Gamma(\mu)}=\frac{\sin(\pi \mu)}{\pi}$.

\noindent
In this regime,
$S(t)$ decreases with time; the decrease in time of the density of events results in the aging of
the correlation function  $\Pi(t_w,t_w+\tau)$, which characteristic time scale is
proportional to the
age $t_w$.

On the other hand, if $\mu > 1$, 
${\hat \psi}(z) = 1 +
\frac{1}{1-\mu} \tau_0 z + o(\tau_0 z)$.
This implies that ${\hat S}(z) \simeq \frac{\mu -1}{\tau_0 z}$, and
that for $t \gg \tau_0$,
$S(t) \simeq \frac{\mu -1}{\tau_0} =\frac{1}{<\tau>}$.
In this non-aging regime, the sprinkling density is uniform in time
and simply equal to the inverse trapping time 
$<\tau>=\int_0^{\infty} d\tau \tau \psi(\tau) =\frac{\tau_0}{\mu -1}$.

\subsection{Definition of the system of study as a superposition of $N$ subsystems}

Let us now turn to our system of interest. This new system is defined as the superposition of
$N$ subsystems, identical to the one introduced previously, each of which is defined by the same
trapping times distribution:
$\psi(\tau)=\frac{\mu \tau_0^{\mu}}{(\tau_0+\tau)^{1+\mu}}$.
The subsystems are assumed to be independent of each other.
One can give an interpretation of such a model in a real space representation: given 
a system of size $L$ in $d$ dimensions, we assume that one can divide this system into
$N$ independent subsystems of length $\xi$, where
$N=(\frac{L}{\xi})^d$. $\xi$ is the typical coherence length of the system, and is
considered constant during the time relaxation. However, we will see that this quantity is
susceptible to evolve during aging.

During the dynamical evolution, each system relaxes independently, and hence contributes to
some extent to the relaxation of the whole system. We will make the following assumptions:
{\it (i)} all events occuring in a subsystem are also defined  as individual events for the whole
system, {\it (ii)} all events contribute equally to the relaxation of the whole system.

This translates into the following definitions:

$\bullet$  {\it (i)} The analog for the whole system
of $\psi(\tau)$  will be denoted as $P_N(\tau)$: it
is the distribution of time intervals between all events (i.e trapping times). We will see in
the next section that this quantity  depends in general on the age $t_w$; we will
then call it $P_N(\tau,t_w)$.

$\bullet$  {\it (ii)} The correlation function of the whole system is defined as
$$C(t_w, t_w+\tau)=\frac{1}{N} \sum_{i=1}^N C_i (t_w, t_w+\tau),$$

\noindent
where $C_i (t_w, t_w+\tau)$ is the correlation function of subsystem $i$.

Before turning to a detailed calculation of $P_N(\tau,t_w)$, we can invoke an argument of
statistics of extremes. If $\tau$ is a trapping time of the whole system, then it seems
natural to say that $\tau={\rm min} \{\tau_i\}_{i=1,..,N}$, where $\tau_i$
is a trapping time of each subsystem $i$. However, this is true only if all subsystems undergo
one event at some time origin, and that one computes the first trapping time of the whole
system from this time origin. Hence this argument definitely excludes aging effects, because
it neglects any memory effects in the dynamics of the subsystems.

Having made this approximation, one can follow a standard calculation of statistics of
extremes, and one can find the distribution of time intervals $P_N(\tau)$:

$$P_N(\tau)=N\psi(\tau) \left[ \int_{\tau} ^{\infty} dt' \psi(t') \right]^{N-1}
= N\mu \frac{\tau_0^{\mu N}}{(\tau_0+\tau)^{1+\mu N}}$$

By expanding around the most probable value $\tau=0$, and setting 
$u=\mu N \tau/ \tau_0$, one finds the limiting  exponential distribution 
$P(u)=e^{-u}$ for $u \ll 1$ (i.e $\tau \ll \frac{\tau_0}{\mu N}$).
This corresponds in fact to the
 convergence of the probability distribution of extremes
towards the Weibull distribution, in the case where the
elementary distribution $\psi(\tau)$ has a finite value
 for its minimum time $\tau=0$ \cite{GUMBEL}.

In this approximation, the time distribution of events of the whole system simply follows a
Poisson process, with a rate proportional to the number of subsystems. In a Poisson process,
the conditions of the experiment are supposed to remain constant in time, and all events are
independent of each other. However, in this model, although single events in all subsystems
are indeed independent of each other, the dynamics is not invariant under time translations
(as can be inferred from the sprinkling density $S(t)$). As we shall see in the next section,
this will give rise to more complicated laws for $P_N(\tau,t_w)$.

Note finally that we have not been able to find a suitable argument of statistics of extremes
for $C(t_w, t_w+\tau)$ (the statistics of this quantity in the framework of non-equilibrium
dynamics have been related to the Gumbel distributions \cite{BRAMWELL, CHAMON}, which are one
of the ``universal'' families of probability distributions of extremes).
Instead, we will gain information (see section V)  by studying $C(t_w, t_w+\tau)$
as the sum of $N$ random variables, reinforcing the idea that it is not an extremal quantity,
but rather originates in the contribution of many individual events (as was already pointed
out in \cite{BRAMWELL}).

\section{III. Distribution of time intervals between all events}

Let $P_N(\tau,t_w)$ be the probability that an event takes place at $t_w+\tau$
if one took place at $t_w$, in
the system composed of $N$ independent subsystems.

In this section, it will be more practical 
for the computation to work with the cumulative probability distribution
$P^C_N(\tau,t_w)=\int_{\tau}^{\infty} d\tau' P_N(\tau',t_w)$.
By definition, it is the probability that the time difference between two successive events is
larger than $\tau$. Similarly, we will use
$Q(\tau)=\int_{\tau}^{\infty} d\tau' \psi(\tau')=
\tau_0^{\mu}/(\tau_0+\tau)^{\mu}$, which is the probability for a trapping
time of a subsystem to be larger than $\tau$, i.e the probability for a subsystem not to
change trap during the period of time $\tau$.

We now call $i$ the subsystem in which
 one event has taken place at $t_w$. 
Then let 
$\{t_j\}_{j \neq i}$ be the  $(N-1)$
times of the last events before
$t_w$ in the other subsystems $j$. The next event to take place in the whole system will
either happen in subsystem $i$ or in any other subsystem. In order for this next event to
occur after a time $\tau$, one requires the following conditions:

$\bullet$ {\it (i)} subsystem $i$ has to remain trapped between $t_w$ and $t_w+\tau$, with
probability $Q(\tau)$, and

$\bullet${\it (ii)} the other subsystems $j$ have to remain trapped between $t_j$ and 
$t_w+\tau$, with probability 
$ Q(t_w+\tau)+ \int_0^{t_w} dt_j S(t_j) Q(t_w+\tau -t_j)$,
$ Q(t_w+\tau)$ being the contribution for the special case $t_j=0$.
This last probability is in fact equal to $\Pi(t_w,t_w+\tau)$, the probability for a system to remain trapped
between times $t_w$ and $t_w+\tau$ (see section II).

\noindent
Hence,

$$P^C_N(\tau,t_w)=  Q(\tau)
 \left[ \Pi(t_w,t_w+\tau) \right]^{N-1}.$$

\noindent

$$P_N(\tau,t_w)= -
\frac{\partial}{\partial \tau} \left[Q(\tau) \left [\Pi(t_w,t_w+\tau)\right ]^{N-1}\right].$$

In the following, we will always consider the case of large $t_w$: $t_w \gg \tau_0$. We now
treat separately  two different regimes for $\tau$:
${\it (i)}$  $\tau \sim \tau_0$, and ${\it (ii)}$  $x=\frac{\tau}{t_w}$ finite and smaller than $1$.

\subsection {Case $\tau \sim \tau_0$}

In the regime of interest where $t_w \gg t_0$,
$S(t_w)\simeq \frac{1}{t_0^{\mu}} \frac{c(\mu)} {t_w^{1-\mu}}$ (see section II). So that the
leading term in $t_w$ is:

$$ \Pi( t_w, t_w+\tau) \simeq 1- \tau  \ S(t_w)  \simeq e^{-\tau S(t_w)} $$

\noindent
Then, we get the result:

\begin{equation}
P^C_N(\tau,t_w)\simeq Q(\tau)  \ 
 e^{-(N-1)\frac{c(\mu)\tau}{\tau_0^{\mu} t_w^{1-\mu}}} \label{case1}
\end{equation}

\noindent
and 

$$ P_N(\tau,t_w)\simeq \left(\psi(\tau) +(N-1) S(t_w)Q(\tau)\right)
e^{- (N-1) \tau S(t_w)}
$$

We note that for $N=1$ the result $P_N(\tau,t_w)=\psi(\tau)$ is recovered.
As $N$ increases, the exponential part in (\ref{case1}) becomes dominant, introducing a rate
that is dependent on the waiting time: $\rho=\frac{c(\mu) (N-1)}{\tau_0^{\mu} t_w^{1-\mu}}$.

This can be interpreted as an effective Poisson process, where the number of instances $N$ is
replaced (for large $N$) by 
$$N_{eff}(t_w)=N \left( \frac{\tau_0}{t_w}\right) ^{1-\mu}.$$

In other words, the scaling of (\ref{case1}) suggests that computing at age $t_w$
the distribution of
events of a system composed initially of $N$ subsystems is the same as computing the
distribution of events of a ``young'' system composed of $N_{eff}(t_w)$ subsystems.
In the case of (\ref{case1}), one can see that $N_{eff}(t_w)$ decreases explicitly with the age;
in the limiting case where aging disappears ($\mu \rightarrow 1$), $N_{eff}$ is simply a
constant equal to $N$. The idea of a number of independent subsystems decreasing with the age
in non-equilibrium systems is not new. It is intimately related to the concept
of a growing lengthscale in an aging system. If one defines a typical length
$\xi$ of a subsystem by $N=(\frac{L}{\xi})^d$, one has
$\xi_{eff}= L N_{eff}^{-1/d}$, and the dependence of $N_{eff}$ on $t_w$ induces the following
power-law  for $\xi_{eff}$:

$$ \xi_{eff}(t_w)= L N^{-1/d} \left(\frac{t_w}{\tau_0}\right)^{\frac{1-\mu}{d}}$$

\subsection {Case $x=\frac{\tau}{t_w} \ll 1$ }

In the regime most accessible experimentally, $x=\frac{\tau}{t_w} \ll 1$, and
$\Pi(t_w,t_w+\tau) \simeq 1 -c(\mu) \left( \frac{\tau}{t_w} \right)^{1-\mu}$, which leads to
the result:

\begin{equation}
P^C_N(\tau,t_w)\simeq  Q(\tau) \ 
 e^{-(N-1) c(\mu) \left(\frac{\tau}{t_w}\right)^{1-\mu}} \label{case2}
\end{equation}

and 

\begin{eqnarray*}
P_N(\tau,t_w) && \simeq  [ \psi(\tau) + \\
&& \frac{(N-1) c(\mu)(1-\mu)}{t_w^{1-\mu}\tau^{\mu}} Q(\tau) ]
 e^{-(N-1)c(\mu) \left(\frac{\tau}{t_w}\right)^{1-\mu}}
 \end{eqnarray*}
 
Again the limiting case $P^C_N(\tau,t_w)=\psi(\tau)$ for $N=1$ is recovered.
In this regime, when $N$ increases, the distribution $P^C_N(\tau,t_w)$
evolves towards a stretched exponential in $\tau$, with a characteristic time proportional to
the age $t_w$.
But as in the case of (\ref{case1}), the scaling of (\ref{case2}) suggests 
that the system at $t_w$ is equivalent to a ``young'' system composed of $N_{eff}(t_w)$ subsystems
with $N_{eff}(t_w)=N \left( \frac{\tau_0}{t_w}\right) ^{1-\mu}$.

Let us now be more precise concerning the relevance of the
quantity $N_{eff}(t_w)$.
 The physical meaning of the effective number of
 subsystems and of the effective coherence length is the following. 
 Consider the system at time $t_1$, with a coherence length $\xi_1$. 
 This system is hence made
 of $N_1$ independent subsystems of size $\xi_1$, by definition of the model; and each
 subsystem is a trap model that can hop in different states in time, with average hopping
 rate $S(t_1)$. 
 Each subsystem ages, so that at some later time $t_2$, the hopping rate decreases
  and is equal to
 $S(t_2)$. This also means that per unit time, less subsystems have hopped
 than at age $t_1$.  In a real space picture, the subsystems that hop are then more sparse
 and far away from each other. This induces some kind of enhanced spatial correlation.
 Then one can make a coarse graining of subsystems and define bigger subsystems, of size
 $\xi_2$, in such a way that the hopping rate of these new subsystems is the same as
 at time $t_1$, which is possible, precisely because they are bigger. Hence the total system 
 at time $t_2$ is now  a collection of $N_2$ subsystems of size $\xi_2$, each of
 which hops at the same rate as the $N_1$ subsystems of size $\xi_1$ at time $t_1$.
 This can actually be quantified by a simple argument. The average hopping rate of one
 subsystem at time $t_1$ is $S(t_1)$. At time $t_2$, one defines a coarse-grained subsystem as
 composed of $N_1/N_2$ of the former subsystems. Then the average hopping rate of one
 coarse-grained subsystem at time $t_2$ is $S(t_2) N_1/N_2$. The hopping rates are 
 chosen to be equal,
 which leads to  $N_2=N_1 S(t_2)/S(t_1)$.
 In the special case of $t_1=\tau_0$ and $t_2=t_w$, one finds:
 $N(t_w)=N (\frac{\tau_0}{t_w})^{1-\mu}$,
 which is exactly the relation  for $N_{eff}(t_w)$ found from the
 previous calculations.

To conclude this section, the study of the two cases investigated above show that for a small
system, $P^C_N(\tau,t_w)$ (and $P_N(\tau,t_w)$) will still be very close to the power-law
characterizing one single subsystem. For a very large system, $P^C_N(\tau,t_w)$
crosses over from an exponential form to a stretched exponential form at larger $\tau$.

In general, there will be a crossover  in $P_N(\tau,t_w)$ from an exponential times a power-law, to a stretched
exponential times a power-law, as $\tau$ increases. In all cases, the distributions become fatter with the age,
which allows to define an effective number of subsystems 
$N_{eff}(t_w)=N \left( \frac{\tau_0}{t_w}\right) ^{1-\mu}$, or equivalently an effective
coherence length
$ \xi_{eff} (t_w)= L N^{-1/d} \left(\frac{t_w}{\tau_0}\right)^{\frac{1-\mu}{d}}$.

For illustration, we  plot on Figure 1 the cumulative probability
distribution $P^C_N(\tau,t_w)$ in the two regimes studied for different values of $N$; the
values of the parameters are $\tau_0=1$, $\mu=0.5$ and $t_w=100$.

\begin{figure}
\includegraphics[width=6cm, height=8cm,angle=270]{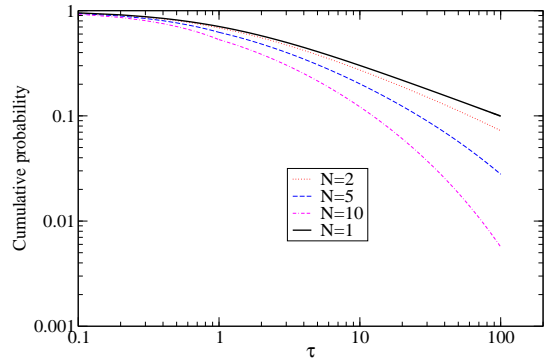}
\caption{(Color online) $P^C_N(\tau,t_w)$ in the two regimes studied for $N=1,2,5,10$; the
values of the parameters are $\tau_0=1$, $\mu=0.5$ and $t_w=100$. As $N$ is increased, the
departure from the power-law is observed. 
\protect\label{timedistrib}}
\end{figure}

\section{IV. Distributions of time intervals between decorrelations. 
Application to experimentally accessible data}

\subsection{1. Definition of the quantities of interest}

When one is not able experimentally to identify
individual rearrangement events, it may be easier to turn to the study of the fluctuations of
global quantities such as correlation functions. More precisely, in the scattering experiments
of \cite{CIP}, non averaged correlation functions called $c_I(t_w, \tau)$ are computed. 
$c_I(t_w, \tau)$ represents the degree of correlation between the speckle field scattered by
the sample at time $t_w$, and the one scattered at time $t_w+\tau$. The time lag $\tau$ can be
given a fixed value during the analysis of the data, and one computes the time series
of $c_I(t_w, \tau)$ as a function of time, starting from time $t_w$ (see for example Figure
3(a) in \cite{CIP}). This allows to compute the probability distribution of $c_I$, 
$P(c_I)$, for a given $\tau$ and a given age $t_w$. This is exactly what we compute in
section V, if we assume that the function $C(t_w, t_w+\tau)$ of our model
can be identified with $c_I(t_w, \tau)$. The main experimental finding, namely that $P(c_I)$
is a negatively skewed distribution for a value $\tau$ small compared to $t_w$ in a
non-equilibrium system (a colloidal gel), is recovered in Section V.

In the analysis of the time series of $c_I(t_w, \tau)$, one can also compute
the distribution of time intervals between significant decorrelations of the system -i.e big
downward jumps of $c_I(t_w, \tau)$-, which is accessible experimentally. Note that these big
jumps do not necessarily correspond  to the same individual ``events'' studied in Section II
and section III, and actually correspond to the superposition of several of them. However these jumps are the only
visible manifestation of the individual ``events'' from an experimental point of view, unless
new techniques allow to visualize in detail and record
the spatial rearrangements of the particles in
real space.

Typically one would like to compute the distribution of time intervals between the smallest
values of $c_I(t_w, \tau)$ -or $C(t_w, t_w+\tau)$ in our model-, a threshold value $C_{th}$
being fixed. One can write that
$C(t_w,t_w+\tau)=1-\frac{n(t_w,t_w+\tau)}{N}$,
where $n(t_w,t_w+\tau)$ is the number of subsystems that changed trap between 
$t_w$ and $t_w+\tau$. In this section, we will keep the notation $N$ for the number of
subsystems, since we will always place ourselves at a given age $t_w$. The influence of
the dependence of the number of subsystems on $t_w$ will be discussed in the conclusion and is not crucial
here.

The threshold $C_{th}$ can be chosen such that one selects only the situations where at least
$k$ subsystems have changed trap between $t_w$ and $t_w+\tau$, so that $C_{th}=1-\frac{k}{N}$,
 and the values of $C(t_w,t_w+\tau)$ considered will be less than $C_{th}$.
The distribution of time intervals between successive jumps of the correlation of this kind
will be denoted
${\cal P}^{(k)} (\tau, t_w, T)$, where $T$ is the time interval variable.

Before considering the general case, we will first focus on two simpler cases:

$\bullet$ {\it (i)} we will first compute ${\cal P}^{(N)} (\tau, t_w, T)$,
which corresponds to the distribution of times between successive decorrelations of
maximum intensity ($k=N$). We will see that the age $t_w$ is not relevant for this quantity,
and it will be denoted ${\cal P}^{(N)} (\tau, T)$.

$\bullet$ {\it (ii)} we will also compute ${\cal P}^{(k,k)} (\tau, t_w, T)$, the distribution
of time intervals between successive decorrelations of the same intensity, i.e which
correspond to the case where
$k$ subsystems have changed trap between $t_w$ and $t_w+\tau$, and 
$k$ subsystems have changed trap between $t_w+T$ and $t_w+ T +\tau$, for the first time since
$t_w$.

\vskip1cm

\subsection{2. Special case of the largest decorrelations}

In this section we want to compute ${\cal P}^{(N)} (\tau, t_w, T)$.
Since we consider only the largest decorrelations,
we say that an event has occured at $t_w$ if
 $C(t_w,t_w+\tau)$ first reaches the value 0 at $t_w$. 
 We define ${\cal P}_0^{(N)} (\tau, t_w, T)$ 
 as the probability per unit time
 $\tau$ that such an event takes place at time $t_w+T$,
  knowing that such an event happened at
 $t_w$.
This quantity will be helpful in all the following in order to calculate
${\cal P}^{(N)} (\tau, t_w, T)$.

In the case of ${\cal P}^{(N)} (\tau, t_w, T)$, the decorrelations in the intervals
$[t_w; t_w+\tau]$ and $[t_w+T; t_w+T+\tau]$ are two successive total decorrelations; 
whereas in the case of
${\cal P}_0^{(N)} (\tau, t_w, T)$, they may not be successive in time. Formally speaking, 
${\cal P}^{(N)} (\tau, t_w, T)$ plays the role of $\psi(T)$ and 
${\cal P}_0^{(N)} (\tau, t_w, T)$ the role of $S(T)$, where $\psi(T)$ and $S(T)$ have been
introduced in Section II.

Throughout the whole section, we will consider the regime where $\tau \ll T$.
In this case, a good approximation is:

$${\cal P}_0^{(N)} (\tau, t_w, T) \simeq \frac{1}{\tau} \left[\tau S(T)\right]^N.$$

Therefore, this quantity does not depend on age, and, as in section II, one has the relation:

$${\cal P}_0^{(N)} (\tau, T)= {\cal P}^{(N)} (\tau, T)
+ \int_0^T dt' {\cal P}^{(N)} (\tau, t') {\cal P}_0^{(N)} (\tau, T-t').$$

Equivalently, if $z$ is the Laplace variable conjugated to $T$, the Laplace transforms are
related according to

$${\hat{\cal P}}^{(N)} (\tau, z)=
\frac{{\hat{\cal P}}_0^{(N)} (\tau, z)}{1+{\hat{\cal P}}_0^{(N)}(\tau, z)}.$$

In the case of interest ($\mu <1$), we use the fact that
$S(T) \simeq c(\mu) \frac{T^{\mu-1}}{\tau_0^{\mu}}$ for
$T \gg \tau_0$, so that

$${\hat{\cal P}}_0^{(N)} (\tau, z) \simeq \frac{1}{\tau}
\left[ \frac{c(\mu) \tau}{\tau_0^{\mu}} \right]^N
\frac{1}{z^x}
\int_{\tau_0 z}^{\infty} du  \ e^{-u} u^{x-1},$$

\noindent
where $x=1-N(1-\mu)$, and we have introduced the lower cut-off $\tau_0$ for the case where
the integral is divergent at the origin.

According to the value of $N$, we will see below through the study of the different cases that
${\cal P}^{(N)} (\tau, T)$ is a power-law with an exponent depending on $N$.

\subsubsection{{\bf (a) Case \ $0 < x <1: N(1-\mu) < 1$}}

In this case, the former integral converges when $\tau_0 z \rightarrow 0$; therefore,

$${\hat{\cal P}}_0^{(N)} (\tau, z) \simeq 
\frac{1}{\Gamma(1-x)} \frac{1}{(T_0 z)^x},$$

\noindent
where we have defined
$T_0^x= \tau \left[ \frac{\tau_0^{\mu}}{c(\mu) \tau}  \right]^N
\frac{1}{\Gamma(x) \Gamma(1-x)}$.
Hence, if $T_0 z \ll 1$,
${\hat{\cal P}}^{(N)} (\tau, z) \simeq 1 -\Gamma(1-x) (T_0 z)^x$, which leads to

$${\cal P}^{(N)} (\tau, T) \simeq \frac{x T_0^x}{(T_0 +T)^{1+x}}.$$

Note that in the special case $N=1$, one has $x=\mu$ and $T_0= \tau_0$,
and one recovers the power-law with the
initial exponent $1+\mu$.

\subsubsection{{\bf (b) Case \ $0 < y <1: 1 < N(1-\mu) < 2$}}

The integral is divergent at the origin in the case where $x<0$,
and instead of $x$, we use 
for convenience $y=-x=N(1-\mu)-1 > 0$.

For $0 < y <1$, or $1 < N(1-\mu) < 2$, we use the following expansion when $\tau_0 z
\rightarrow 0$:

$$\int_{\tau_0 z}^{\infty} du \ \frac{e^{-u}}{u^{1+y}} \simeq
\frac{1}{y} \frac{1}{(\tau_0 z)^y} -\frac{1}{y} \Gamma(1-y) $$

We define the time $T_1$ such that 
$T_1^y=\frac{1}{\tau} \left[\frac{c(\mu) \tau}{\tau_0^{\mu}}\right]^N$,
and we find that  
${\hat{\cal P}}_0^{(N)} (\tau, z) \simeq
\frac{1}{y} \left(\frac{T_1}{\tau_0}\right)^y 
-\frac{1}{y} \Gamma(1-y) (T_1 z)^y$, and

$$ {\hat{\cal P}}^{(N)} (\tau, z) \simeq
\frac{\frac{1}{y} \left(\frac{T_1}{\tau_0}\right)^y}
{1+\frac{1}{y} \left(\frac{T_1}{\tau_0}\right)^y}
\left[1- 
\frac{\Gamma(1-y)}{1+\frac{1}{y} \left(\frac{T_1}{\tau_0}\right)^y} (\tau_0 z)^y \right].$$

Hence, ${\cal P}^{(N)} (\tau, T)$ is again a power-law at large $T$:

$${\cal P}^{(N)} (\tau, T) \simeq 
\frac{\frac{1}{y} \left(\frac{T_1}{\tau_0}\right)^y}
{1+\frac{1}{y} \left(\frac{T_1}{\tau_0}\right)^y}
\frac{y T_2^y}{(T_2 +T)^{1+y}},$$

where $T_2= 
\frac{\tau_0}{\left[ 1+\frac{1}{y} \left(\frac{T_1}{\tau_0}\right)^y \right]^{1/y}}$.

Note that the constant in front of the power-law is not exact (one would need the expression
for all $T$ of $S(T)$ to get its exact expression). Moreover, this constant is smaller than
$1$, which means that the distribution ${\cal P}^{(N)} (\tau, T)$
is not normalized. This comes from the fact that the total number of events (which is equal
to ${\hat{\cal P}}_0^{(N)} (\tau, z=0)$) is finite; therefore there is a non-zero
probability that the time interval between two events is infinite, so that 
${\cal P}^{(N)} (\tau, T)$ is not normalized to unity.

\subsubsection{{\bf (c) Case \ $y> 1: N(1-\mu) > 2$}}

Now the expansion for $\tau_0 z
\rightarrow 0$ of the integral reads:

$$\int_{\tau_0 z}^{\infty} du \ \frac{e^{-u}}{u^{1+y}} \simeq
\frac{1}{y} \frac{1}{(\tau_0 z)^y} -\frac{1}{y-1}\frac{1}{(\tau_0 z)^{y-1}}
+\frac{1}{y}\frac{1}{1-y} \Gamma(2-y).$$

in the case where $1<y<2$. In general, if $n=E(y)$ is the integer part of $y$, the constant part of the expansion is
proportional to $\Gamma(n+1-y)$ and is followed by a $(\tau_0 z)^{n+1-y}$ term.
Then, following the case $1<y<2$, we find that
${\hat{\cal P}}_0^{(N)} (\tau, z) \simeq
\frac{1}{y} \left(\frac{T_1}{\tau_0}\right)^y 
-\frac{1}{y-1} T_1^y \tau_0^{1-y} z  
+ \frac{1}{y}\frac{1}{1-y} \Gamma(2-y) (T_1 z)^y + o(z^2)$. In general, there will always be a singular term in
$z^y$ in between two polynomial terms $z^n$ and  $z^{n+1}$.
Again in the special case where $1<y<2$, we find: 
$ {\hat{\cal P}}^{(N)} (\tau, z) \simeq
\frac{\frac{1}{y} \left(\frac{T_1}{\tau_0}\right)^y }
{1+\frac{1}{y} \left(\frac{T_1}{\tau_0}\right)^y}
(1-\frac{1}{y-1} y T_2^y \tau_0^{1-y} z + \frac{1}{y-1}
\Gamma(2-y) (T_2 z)^y +o(z^2))$.

As a reminder, let us note that the Laplace transform for small $z$ of a power-law distribution $\psi(\tau)$ introduced in Section
II is, for $1<\mu<2$:
$\hat{\psi}(z) = 1 -\frac{1}{\mu-1} \tau_0 z +\frac{1}{\mu-1} \Gamma(2-\mu) (\tau_0 z )^{\mu} + o((\tau_0 z)^2)$.
Hence $ {\hat{\cal P}}^{(N)} (\tau, z)$ cannot be mapped exactly on this type of function, and one can for example
add an exponential function to a power-law in order to recover the correct  expansion up to order $z^2$. This
introduces some indeterminacy in the determination of ${\cal P}^{(N)} (\tau, T)$ .
However, the behaviour at very large $T$ will be:

$${\cal P}^{(N)} (\tau, T) \simeq 
\frac{\alpha}{T^{1+y}},$$

where $\alpha$ is an indetermined coefficient. This final power-law behaviour might be in fact rather hard to
observe in reality.

\subsection{3. Special case of the decorrelations with same value}

In this section we compute the quantity
${\cal P}^{(k,k)} (\tau, t_w, T)$.
Like for the previous case, we define ${\cal P}_0^{(k,k)} (\tau, t_w, T)$
corresponding to ${\cal P}^{(k,k)} (\tau, t_w, T)$, which measures the probability that $k$
subsystems change traps in the interval $[t_w+T;t_w+T+\tau]$, knowing that $k$ subsystems
(not necessarily the same)
have changed traps in the interval $[t_w;t_w+\tau]$.
More precisely, in this section, we say that an event has occured at $t_w$ if
$C(t_w,t_w+\tau)$ first reaches the value $1-\frac{k}{N}$ at $t_w$. 
Then ${\cal P}_0^{(N)} (\tau, t_w, T)$ 
 is the probability per unit time
 $\tau$ that such an event takes place at time $t_w+T$,
  knowing that  an event happened at
 $t_w$.
In this case ($k\neq N$), all quantities will depend on $t_w$ and a relation of the type of
(\ref{sprinkle}) is a priori not correct. However, in the limit that we consider, where both $T$ and
$\tau$ are small compared to $t_w$, and if $k$ is not too far from $N$ (experimentally, the
threshold will be taken low enough in order to get rid of experimental noise), we will assume
that such a relation can still be valid. We will write:

\begin{eqnarray}
& &{\cal P}_0^{(k,k)} (\tau, t_w, T) ={\cal P}^{(k,k)} (\tau, t_w, T) \nonumber\\
&+ &\int_0^T dt' {\cal P}^{(k,k)} (\tau, t_w, t') {\cal P}_0^{(k,k)} (\tau, t_w, T-t') \label{sprinkle_k}
\end{eqnarray}

Like in the previous section, we will first compute
${\cal P}_0^{(k,k)} (\tau, t_w, T)$ and deduce ${\cal P}^{(k,k)} (\tau, t_w, T)$
from its Laplace transform thanks to the relation:

\begin{equation}
{\hat{\cal P}}^{(k,k)} (\tau, t_w, z)=
\frac{{\hat{\cal P}}_0^{(k,k)} (\tau, t_w, z)}{1+{\hat{\cal P}}_0^{(k,k)}(\tau, t_w, z)} \label{laplace_k}
\end{equation}

In order to compute ${\cal P}_0^{(k,k)} (\tau, t_w, T)$, we call $n$ the number of subsystems
that changed traps in the interval $[t_w, t_w+\tau]$, but stayed trapped in the interval
$[t_w+T, t_w+T+\tau]$. Hence $k-n$ subsystems change traps both in the intervals
$[t_w, t_w+\tau]$ and  $[t_w+T, t_w+T+\tau]$; $n$ subsystems change traps in the interval 
$[t_w+T, t_w+T+\tau]$ but where trapped in the interval $[t_w, t_w+\tau]$; finally,
$N-k-n$ subsystems were trapped in both the intervals 
$[t_w, t_w+\tau]$ and  $[t_w+T, t_w+T+\tau]$.

Therefore, if $k_{sup}={\rm min}(k,N-k)$,

\begin{eqnarray*}
 & & {\cal P}_0^{(k,k)} (\tau, t_w, T)= 
 \frac{C_N^k}{\tau} \sum_{n=0}^{k_{sup}}
C_k^n C_{N-k}^n \left[ \tau S(T) \right]^{k-n} \times\\
& & \left[ 1-\tau S(T) \right]^{n}
\left[ \tau S(t_w+T) \right]^{n}
\left[ 1-\tau S(t_w+T) \right]^{N-k-n}.\\
\end{eqnarray*}

Such a finite sum being analytically untractable, we will take its continuous limit, by
defining $x=k/N$ and $y=n/N$, and by replacing $\sum_{n=0}^{k_{sup}}$
by $N \int_0^{x_{sup}}$ in the limit where $N$ is large.

First, if we consider the lowest order in $\tau$ only, we reduce the expression to:

$${\cal P}_0^{(k,k)} (\tau, t_w, T)= 
 \frac{C_N^k}{\tau} \sum_{n=0}^{k_{sup}}
C_k^n C_{N-k}^n \left[ \tau S(T) \right]^{k-n} 
\left[ \tau S(t_w+T) \right]^{n}.$$

Then by using the Stirling formula $N! \sim N^N e^{-N} \sqrt{2\pi N}$, and taking the
continuous limit, we get:

\begin{eqnarray}
& &{\cal P}_0^{(k,k)} (\tau, t_w, T)\simeq 
\frac{\tau^{k-1}}{(2\pi)^{3/2} N^{1/2}}
\int_0^{x_{sup}} dy \nonumber \\
& & \exp [-N
\{(x-y) \ln(x-y) \nonumber \\
&+& 2y \ln y +(1-x-y) \ln (1-x-y) \nonumber \\
&-& (x-y) \ln S(T) -y \ln S(T+t_w) \} \label{p0kk}] 
\end{eqnarray}

By using a saddle point method, we find that the  former integral is maximized for $y^*$
such that

\begin{equation}
(x-y^*)(1-x-y^*)=y^{*2} \gamma(t_w,T) \label{eq_y*}
\end{equation}
where $\gamma(t_w,T)=\frac{S(T)}{S(T+t_w)}$.

We keep the positive solution, having checked that we have both $y^* \leq x$ and $y^* \leq
1-x$:

$$y^*=\frac{1}{2(\gamma -1)}
[-1+ \sqrt{1+4x(1-x)(\gamma -1)}].$$

Replacing $y^*$ in the last expression for
${\cal P}_0^{(k,k)} (\tau, t_w, T)$ leads to a complicated expression that can be simplified
in our regime of interest,
$\tau_0 \ll T \ll t_w$.

First, by using the expression for $S(T)$ at large times and $\mu < 1$, we find that, for
$\tau_0 \ll T \ll t_w$, $\gamma(t_w,T) \simeq \left( t_w/T \right)^{1-\mu}$, and
$y^*(T, t_w) \simeq \left( \frac{x(1-x)}{\gamma(t_w,T)} \right)^{1/2}$.
Then, if $\frac{t_w}{T}$ is large enough, we can always assume to have
$y^* \ll {\rm min}(x,1-x)$, so that equation (\ref{p0kk}) can be expanded in $y^*$. Finally, we find the
result:

$${\cal P}_0^{(k,k)} (\tau, t_w, T) \simeq \frac{C_N^k}{ 2 \pi \tau}
\left[\tau S(T)\right]^k 
e^{ 2 \sqrt{k(N-k)} \left(\frac{T}{t_w}\right)^{\frac{1-\mu}{2}} }.$$

The Laplace transform with respect to $T$ reads:

$${\hat{\cal P}}_0^{(k,k)} (\tau, t_w, z) \simeq \frac{C_N^k}{ 2 \pi \tau}
\left[ \frac{c(\mu) \tau}{\tau_0^{\mu}}\right]^k  I(t_w,z),$$

where 

$$I(t_w,z)= \int_0^{\infty} dT \ \frac{e^{-zT}}{T^{(1-\mu)k}} 
 \exp  \left[ K(k)  \left(\frac{T}{t_w}\right)^{\frac{1-\mu}{2}} \right],$$
 
 and $K(k)=2 \sqrt{k(N-k)}$.

$I(t_w,z)$ is a convergent integral for $k(1-\mu) < 1$, which is the first case we consider.

\subsubsection{{\bf (a) Case \ $k(1-\mu) <1$}}

Since we are interested in the case $T \ll t_w$, we will make an expansion for $zt_w \gg 1$.

Then we need to consider two separate cases for the evaluation of 
${\hat{\cal P}}^{(k,k)} (\tau, t_w, z)$ from relation (\ref{laplace_k}).
We report here the main results, and refer to Appendix A for the calculations.

\subsubsection{{\bf (a1) Case \ $0 < k < \frac{1+\mu}{2(1-\mu)}$}}

${\cal P}^{(k,k)} (\tau, t_w, T)$ is a power-law:

\begin{equation}
{\cal P}^{(k,k)} (\tau, t_w, T) \simeq 
\frac{C_N^k}{ 2 \pi \tau}
\left[ \frac{c(\mu) \tau}{\tau_0^{\mu}}\right]^k
\frac{A t_w^{\alpha}}{\Gamma(-\beta)} \frac{1}{T^{1+\beta}} \label{power}
\end{equation}

with 
$\alpha=\frac{k(1-\mu)^2}{1+\mu}$ and 
$\beta= \frac{2k(1-\mu)}{1+\mu} -1$.

\subsubsection{{\bf (a2) Case \ $ \frac{1+\mu}{2(1-\mu)}< k < \frac{1}{1-\mu}$}}

\begin{equation}
{\cal P}^{(k,k)} (\tau, t_w, T) \simeq  
e^{\cos (\frac{\pi}{1+\beta} )
(\beta {\cal K}(t_w,\tau))^{\frac{1}{1+\beta}} 
\left(\frac{T}{t_w}\right)^{\frac{\beta}{1+\beta}}
}\label{stretched}
\end{equation}

which is a stretched exponential since $\cos ( \frac{\pi}{1+\beta}) <0$.

\subsubsection{{\bf (b) Case \ $k(1-\mu) >1$}}

${\cal P}^{(k,k)} (\tau, t_w, T)$ is simply an exponential, with a weak dependence on $t_w$
(which vanishes for large $t_w$):

\begin{equation}
{\cal P}^{(k,k)} (\tau, t_w, T) \simeq 
{\cal C}(t_w,\tau)  e^{-{\cal C}(t_w,\tau) T} \label{exp}
\end{equation}

 where
${\cal C}(t_w,\tau)=\frac{C_N^k}{ 2 \pi \tau}
\left[ \frac{c(\mu) \tau}{\tau_0^{\mu}}\right]^k
\frac{e^{K(k)\left(\frac{\tau_0}{t_w}\right)^{\frac{1-\mu}{2}}}}{\tau_0^{(1-\mu)k}}$.

We note here that 
it is not possible to make a clear
 continuity between the cases $k=N$ and the case $k$ different than $N$ treated in this
 section.
 This is due to the face that we use a
 saddle-point approximation of $I(t_w,z)$, which is defined only if $K(k)$ is different from
 $0$, hence $k$ is different from $N$ and $0$. In this section, the approximations 
 are more numerous and
 are susceptible to be valid only for some intermediate values of k. Moreover,
  we are interested in the aging regime and now the correct
 expansion is to be made for $zt_w  \gg 1$, and not anymore for $z \tau_0 \ll 1$. 
 
\subsection{4. Generalization}

The most general case, as introduced in the beginning of Section IV, consists in computing
${\cal P}^{(k)} (\tau, t_w, T)$ the distribution of time intervals between successive
decorrelations larger than $\frac{k}{N}$.

As before, we will use the ${\cal P}_0$ quantities relative to the 
${\cal P}$ quantities we are looking for, and we will assume that relations between them
similar to equation (\ref{sprinkle_k}) still hold.

Therefore, we will start by computing

\begin{equation}
{\cal P}_0^{(k)} (\tau, t_w, T)= \sum_{k' \geq k}  \sum_{k'' \geq k}
{\cal P}^{(k',k'')}_0 (\tau, t_w, T) \label{sum}
\end{equation}

\noindent
where ${\cal P}^{(k',k'')}_0 (\tau, t_w, T)$ is the probability that $k'$ subsystems have
changed traps in the interval
$[t_w,t_w+\tau]$ and $k''$ subsystems have changed traps in the
interval $[t_w+T; t_w+T+\tau]$.
We refer the reader to Appendix B for technical details of the computation, which is similar to the one explained in
section IV.3.

From the calculation, we get

$${\cal P}_0^{(k)} (\tau, t_w, T)={\cal P}_0^{(k^*)} (\tau, t_w, T),$$

$${\cal P}^{(k)} (\tau, t_w, T)={\cal P}^{(k^*)} (\tau, t_w, T),$$

\noindent
where $k^*$ is a growing function of $k$, and is equal to $k$ if  $\tau S(T)< 1$ and $k> N/2$.

Therefore the results of the previous section can be used. We conclude that, according to the
value of $k^*$ (which will be equal to $k$ in most cases of interest), 
${\cal P}^{(k)} (\tau, t_w, T)$ can either take a power-law form (equation(\ref{power})), 
a stretched exponential form (equation (\ref{stretched})),
or an exponential form (equation (\ref{exp})), where $k$ has to be replaced by $k^*$.

\section{V. Probability distribution of the total correlator}

In section III, we used the notion of ``events'',
 and calculated their time
distribution, because it was the most natural quantity to compute in the framework of the
model presented here. However, it may be difficult in an experiment  to
identify such events, or even to find a reasonable definition of individual events.

A well-defined quantity that is more readily accessible is the distribution probability of a
correlator $C(t_w,t_w+\tau)$, as has been recently investigated in numerical simulations
\cite{CHAMON}.

In our model, each subsystem $i$ is a two-level system, where the correlation 
$C_i(t_w,t_w+\tau)$ can only take the values $0$ or $1$. More precisely, the probability
distribution of the correlation of a subsystem $i$ is:

$$P(C_i)=f \delta(C_i-1) +(1-f) \delta(C_i),$$

\noindent
where the parameter $f$ coincides with the average value of the correlation for the 
values of  $t_w$ and $\tau$ considered:
$f=\Pi(t_w,t_w+\tau)=<C_i(t_w,t_w+\tau)>$. In the following we will omit temporarily the
dependence of $f$ on $t_w$ and $\tau$, for simplicity in the notations.

For the whole system (the superposition of the $N$ independent subsystems),
 the definition of $C(t_w,t_w+\tau)$ has been given in Section II. One can
actually rewrite this quantity as:

$$C(t_w,t_w+\tau)=\frac{m(t_w,t_w+\tau)}{N}=1-\frac{n(t_w,t_w+\tau)}{N},$$

\noindent
where $m(t_w,t_w+\tau)$ is the number of subsystems that remained in the same energy trap
between $t_w$ and $t_w+\tau$, and $n(t_w,t_w+\tau)$ the number of
subsystems that changed trap between 
$t_w$ and $t_w+\tau$.

Then, in order to find the value $C$ for the correlation between
times $t_w$ and $t_w+\tau$, one has to draw $m$ subsystems among
$N$ instances that are trapped during this time interval, with probability $f$, and $N-m$
subsystems that changed trap during this time interval, with probability $1-f$. Hence, the
probability distribution of $C$, $P(C)$,
is simply the binomial distribution: $P(C)=N P(m)$, with

$$ P(m)=C_N^m f^m (1-f)^{N-m},$$
	  
and $C_N^m$ is the binomial coefficient.

Note that in the limit of a large number of subsystems $N$, the central limit 
theorem holds and the
limiting distribution is a Gaussian:

$$P(C) \rightarrow \sqrt{\frac{N}{2\pi f(1-f)}}
\exp{\left[-\frac{N (C-f)^2}{2 f (1-f)} \right]}.$$

We show on Figure 2 the distributions $P(C)$ 
for different values of the average
correlation: $<C> =0.1, 0.5,  0.9$. 
The full lines correspond to $N=10$ and the dotted
lines to $N=100$, showing a closer resemblance to a Gaussian distribution for $N=100$.

\begin{figure}
\includegraphics[width=6cm, height=8cm,angle=270]{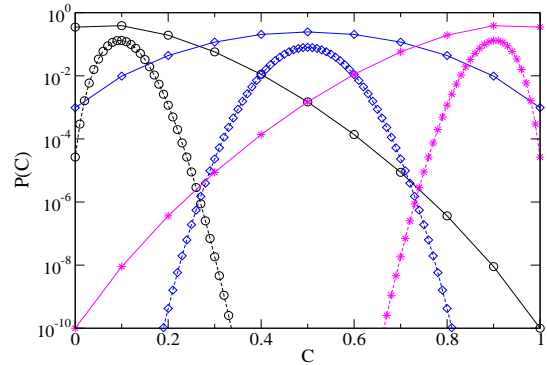}
\caption{(Color online) Probability distribution $P(C)$ for different values of the average
correlation: $<C>=0.1$ (circles), $0.5$ (diamonds), $0.9$ (stars). The full lines correspond to $N=10$ and the dotted
lines to $N=100$.  
\protect\label{proba}}
\end{figure}

For comparison with experiments or numerical simulations, we computed the variance
$\sigma^2(C)$, the skewness $s(C)$ and the kurtosis $\kappa(C)$ of the distribution $P(C)$.
They can be easily computed  by noticing that the cumulants of independent random variables
are additive quantities:

\begin{eqnarray*}
\sigma^2(C)&=&\frac{1}{N} f(1-f)\\
 s(C)&=&\frac{1}{\sqrt{N}}
  \frac{1-2f}{\sqrt{f(1-f)}}\\
\kappa(C)&=&\frac{1}{N} \frac{6f^2-6f+1}{f(1-f)}\\
\end{eqnarray*}

At this stage, one would like to plot the previous quantities as a function of $\tau$ for
different values of the age $t_w$, in order to compare with existing results in the
literature \cite{CIP,CHAMON,FOAMS}.
For given $\tau$ and $t_w$, $f=\Pi(t_w,t_w+\tau)$ is explicitly defined using the relation of
section II. Moreover, we use the results of Section III  that suggest that at age $t_w$, the
system can actually be considered as a superposition of 
$N_{eff}(t_w)= N \left(\frac{\tau_0}{t_w}\right)^{1-\mu}$ independent subsystems. Hence in the
preceding formulae for $\sigma^2(C)$, $s(C)$, $\kappa(C)$, we replace $N$ by $N_{eff}(t_w)$.
We want to stress  that it is not
 proved here that this mapping can be applied to the case of the full probability of the
 correlation functions, it is only a
 phenomenological attempt to try to extend this coarse-graining
 to the calculation of
 other quantities than the distribution of times. This argument actually leads to results
 compatible to what is found in experiments.

In Figures 3, 4 and 5, $\sigma^2(C)$, $s(C)$, and $\kappa(C)$ are shown as a
function of $\tau$
for different values of $t_w$, for $\mu=0.8$ and $N=10$. As observed in recent experiments
on foams \cite{CIP,FOAMS}, one can see that the variance is maximum at a time $\tau$ of the order of the
relaxation time (which in our model is proportional to the age $t_w$), 
and goes to 0 at large $\tau$; this maximum increases with the age. The skewness is
negative at small $\tau$, crosses the origin at intermediate times when $P(C)$ becomes
symetric and then becomes positive; the negative skewness is more and more pronounced when
$t_w$ increases. The kurtosis is positive at small times, then negative, and positive again at
large $\tau$, while the negative part becomes more and more pronounced with the age.
All these results are  compatible with recent numerical simulations \cite{CHAMON}, and
seem to be compatible with preliminary experiments on colloidal gels \cite{CIP}.
Note again that the use of $N_{eff}(t_w)$  contains crucial information: if $N$ was kept constant,
the curves would simply be superimposed by a $\tau/t_w$ rescaling, in particular, the maxima
and minima of the variance, skewness and kurtosis would not depend on the age.

Finally, we computed the probability distribution $P(C)$ for a given $\tau/t_w$ ratio - i.e a
given value of the average correlation $<C>$-, at different ages, therefore for different
$N_{eff}(t_w)$; the system is made initially of $N=1000$ independent subsystems. Figure 6
shows such probability distributions, for $\mu=0.8$, $f=<C>=0.8$, $N=1000$, plotting $P(C)
\sigma(C)$ versus $(C-<C>)/\sigma(C)$. One can
see that even for $N=1000$, systematic deviations exist, though they may be hard to
distinguish when an experimental noise is present. Deviations are systematic as $t_w$
increases, which is not surprising since $N_{eff}(t_w)$ decreases with the age; the central
limit theorem is less and less valid as $t_w$ increases.
Note that we tried to fit these curves with Gumbel distributions
like in \cite{CHAMON}, but we didn't find a very precise agreement
in our case.

Finally, let us note that our results fail to reproduce the expected zero limit of the
skewness and kurtosis of $P(C)$ when $\frac{\tau}{t_w} \rightarrow \infty$. These two quantities
 actually
diverge in our case for any finite $N$, because $P(C)$ tends to 
a delta function $\delta(C)$ and $\sigma(C)
\rightarrow 0$. However, the limit $\frac{\tau}{t_w} \rightarrow \infty$
is formally equivalent to $N_{eff}(t_w) \rightarrow \infty$, which restores the correct limit
because of the central limit theorem.
This problem would be absent if $P(C_i)$ for one individual system had a non-zero variance when $C_i
\rightarrow 0$. In practice, this will be always the case, because of the presence of
white thermal noise.

\begin{figure}
\includegraphics[width=6cm, height=8cm,angle=270]{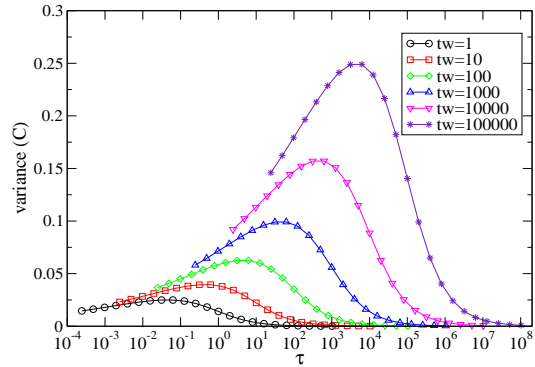}
\caption{(Color online) Variance of the probability distribution $P(C)$ for $\mu=0.8$ and $N=10$ independent
subsystems initially, for different values of $t_w$. 
\protect\label{variance}}
\end{figure}

\begin{figure}
\includegraphics[width=6cm, height=8cm,angle=270]{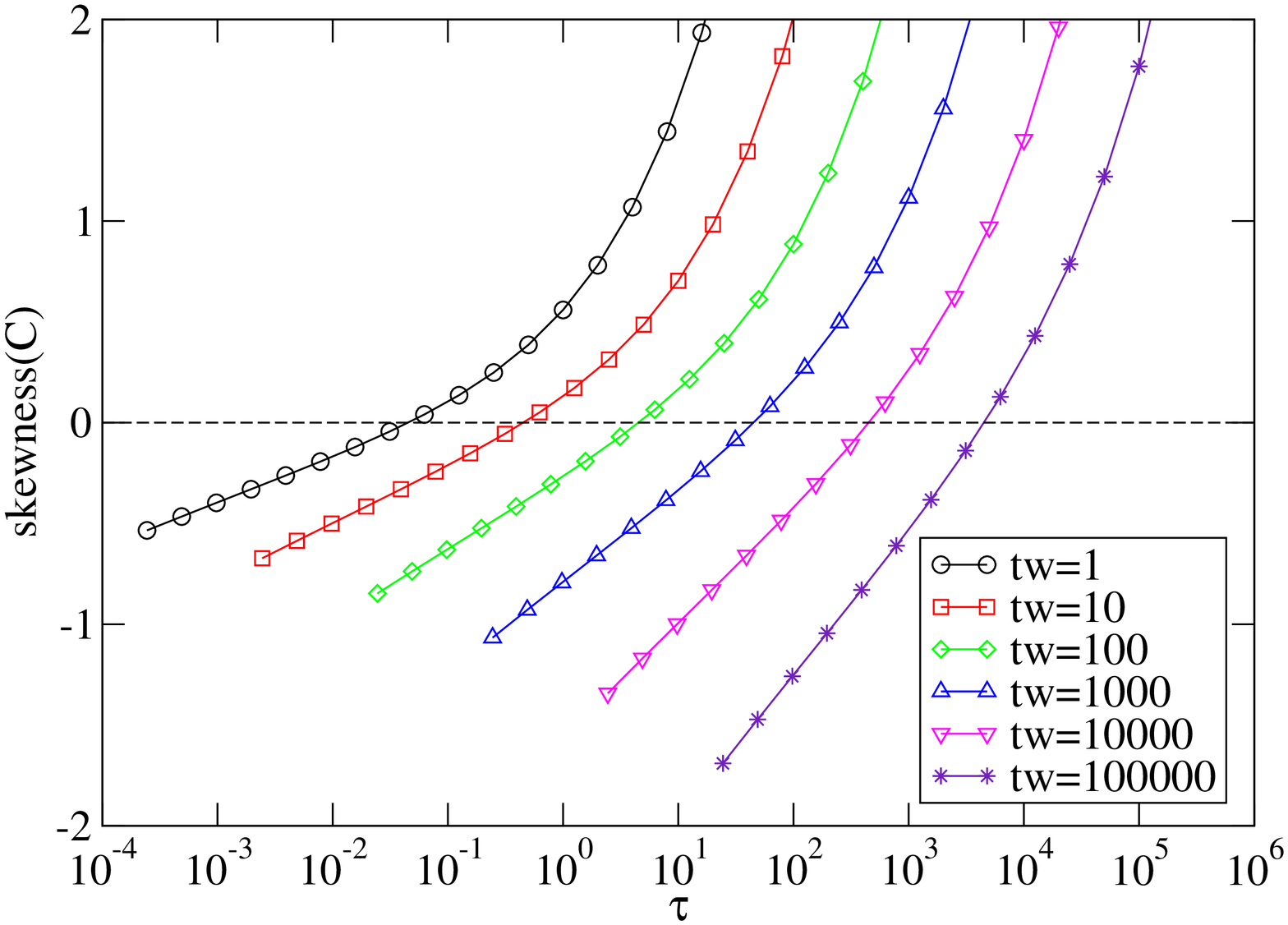}
\caption{(Color online) Skewness of the probability distribution $P(C)$ for $\mu=0.8$ and $N=10$ independent
subsystems initially, for different values of $t_w$.   
\protect\label{skewness}}
\end{figure}

\begin{figure}
\includegraphics[width=6cm, height=8cm,angle=270]{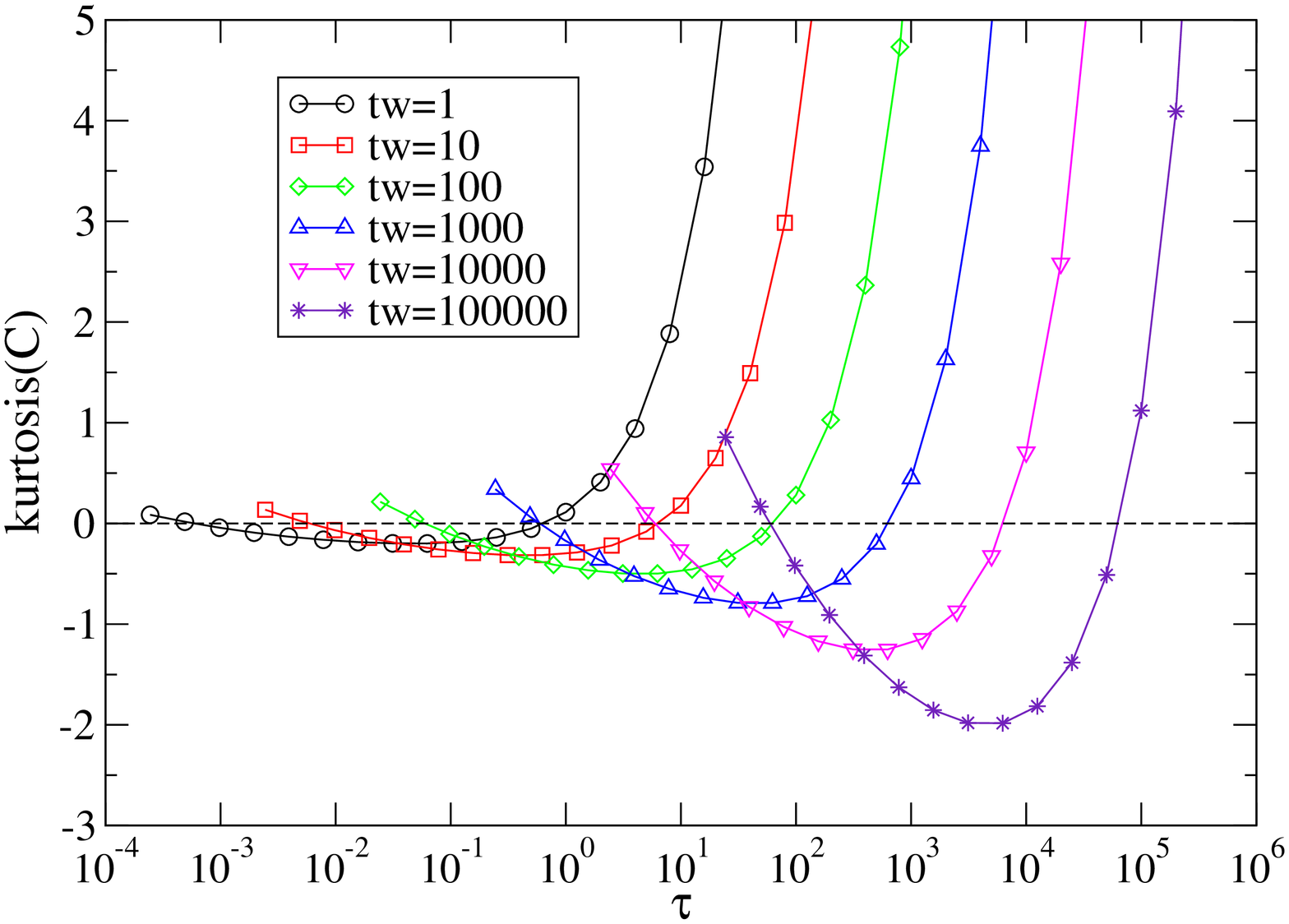}
\caption{(Color online) Kurtosis of the probability distribution $P(C)$ for $\mu=0.8$ and $N=10$ independent
subsystems initially, for different values of $t_w$.   
\protect\label{kurtosis}}
\end{figure}

\begin{figure}
\includegraphics[width=6cm, height=8cm,angle=270]{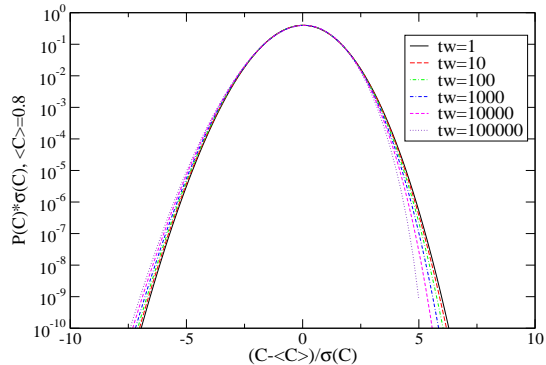}
\caption{(Color online) Rescaled probability distributions $P(C)$ for $\mu=0.8$, $N=1000$ independent
subsystems initially, $<C>=0.8$, and different values of $t_w$.   
\protect\label{rescaling}}
\end{figure}

\section{VI. Discussion and conclusion}

We have studied a system made of the superposition of $N$ subsystems characterized 
individually by a slow
dynamics and a power-law distribution of times between jumps.
This system can in turn be characterized by its distribution of time intervals between
events $P_N(\tau,t_w)$.
For small $N$, the power-law behaviour of a single subsystem is the dominant behaviour.
For large $N$, at very small time intervals, this distribution is exponential and crosses over at
 larger time intervals
to a stretched exponential. In the two regimes, a natural scaling between $N$
and $t_w$ appears, which allows to define an age dependent effective number of subsystems $N_{eff}(t_w)$
which decreases  with $t_w$, or
equivalently, an effective coherence length which grows slowly
with the age:
$\xi(t_w) \sim t_w^{(1-\mu)/d}$.
It would be interesting to see whether experimental results also display the different regimes calculated here. In
particular, one could observe an exponential or a stretched exponential distribution for a young system, and a power-law
distribution for an aged system.
As explained in section III, the interpretation of $\xi_{eff}(t_w)$ comes from the aging of
all individual subsystems: the systems that hop are more rare when $t_w$ increases, and
hence further apart from each other; this introduces a relevant lengthscale $\xi(t_w)$ that
grows with $t_w$.

We also computed
${\cal P}^{(k)} (\tau, t_w, T)$ the distribution of time intervals between successive
decorrelations over an interval $\tau$, larger than $\frac{k}{N}$. This quantity, like $P(C)$,
is accessible through scattering experiments. 
Although the derivation of this quantity is a bit technical, the
results  can be summarized in the following way.
In the extreme case where one only counts the largest decorrelations ($k=N$), the distribution of time intervals
between successive decorrelations crosses over from a power law 
with a small exponent at small $N$ (or large $t_w$) to a
power-law with a large exponent at large $N$ (or small $t_w$).
It is hoped that in experiments, these regimes will be observed, when one varies the age or other
parameters. For example, if the temperature is decreased in a glass, or the density is increased in a gel, or if one
increases the quench rate, one expects that the coherence length increases, i.e the number of independent subsystems
decreases. 
For $k \neq N$, the results are more difficult to obtain, due to technical complications,
and are likely to be reliable only for intermediate values of $k$.
If $k$ is small (i.e one counts
only the decorrelations corresponding to a few subsystems), the distribution 
${\cal P}^{(k)} (\tau, t_w, T)$ is a power-law; if $k$ is larger, it is a stretched exponential and for $k$
large it is an exponential. Equivalently, since one expects the number of subsystems to decrease with the
age, ${\cal P}^{(k)} (\tau, t_w, T)$ is typically
an exponential at small $t_w$,
then a stretched exponential and finally a power-law at large $t_w$,
 for a given value of the correlation threshold $C_{th}=1-k/N$.

Finally, we also computed analytically the probability distribution
of the correlator $P(C)$, which reproduces many features of the experimental findings of \cite{CIP} and of
the numerical results of \cite{CHAMON}.
We used the fact (not shown rigorously) that at $t_w$ the system can be described as a
collection of $N_{eff}(t_w)$ independent subsystems.  No rescaling to some time-evolving Gumbell distribution
seems to be
relevant here. Since the full probability is calculated here, it would be interesting to compare it more quantitatively to
experimental results. In particular, we find that the variance grows with the age like $\sigma^2 \sim N^{-1}_{eff}
\sim t_w^{1-\mu}$.

Although these results are derived
 from  a simplified theoretical model, we think that such a
description captures the essential intermittent character of the relaxation of a realistic glassy system,
together with the existence of a coherence lengthscale growing with the age,
due to the decrease of the rate of events in time.
We hope that these findings will be helpful in analyzing more quantitatively some experimental results. In
particular, the results of Section IV and Section V might be of
interest for the Time Resolved Correlation experiments of \cite{CIP}. However, since the individual trap
model is probably a very crude model for the relaxation of realistic materials ($<C(t_w, t_w+\tau>$ is a
priori unlikely to coincide with the experimental average relaxation function 
of colloidal gels), it may be interesting in a further study to replace it with a more realistic dynamical
model for jammed colloidal gels \cite{gel}.
Concerning the intermittency dynamics studied in \cite{CILIBERTO}, the quantities calculated in Section III
are maybe the most relevant ones to compare with the experimental results.

The model studied here has the major disadvantage that the relevant coherence length of the system
is introduced by hand through the superposition of several subsystems. However, we were able to extract
useful information from this picture. It would be  more satisfactory to solve a model where an aging
coherence length builds up during the dynamics, resulting from microscopic dynamics. Candidates could be
special cases of kinetically constrained models, though analytical calculations will be quite hard
\cite{WHITELAM}. However, numerical simulations  provide now a useful 
alternative tool for the precise investigation, in microscopic models of glasses, of the quantities calculated
in this paper \cite{CHAMON}.

\vskip 1cm

{\it Acknowledgements}$\ \ $  
I would like to thank J-P. Bouchaud for his collaboration at the beginning of this
work. I am particularly indebted to H. Bissig, L. Cipelletti, L. Ramos and V. Trappe
for numerous and fruitful discussions on related experiments. L. Buisson and
S. Ciliberto have to be acknowledged  for useful discussions. I would like to thank L. Berthier and W. Kob for useful
comments and suggestions on this paper. This work has been
supported in part by the European Community's Human Potential Program under contract
HPRN-CT-2002-00307  ``DYGLAGEMEM'', 
the European Community's
Marie Curie Research and Training Network 
on Dynamical Arrest of Soft Matter and Colloids: MRTN-CT-2003-504712,
and ACI "Jeunes chercheuses et jeunes chercheurs" JC2076.

\section {APPENDIX A}

\subsubsection{{\bf (a) Case \ $k(1-\mu) <1$}}

In this case, we can perform a saddle-point calculation in order to evaluate
$I(t_w,z)$, which leads to:

$$I(t_w,z) \simeq A t_w^{\alpha}  z^{\beta} 
 e^{-\frac{B}{(zt_w)^{\gamma}}},$$
 
\noindent
with 
$A=[(1-\mu)K(k)/2]^{-\frac{2k(1-\mu)}{1+\mu}}$,
$B=[(1-\mu)K(k)/2]^{\frac{2}{1+\mu}}$,
$\alpha=\frac{k(1-\mu)^2}{1+\mu}$,
$\beta= \frac{2k(1-\mu)}{1+\mu} -1$, and $\gamma=\frac{1-\mu}{1+\mu}$.

Since we are interested in the case $T \ll t_w$, we expand $I(t_w,z)$ for $zt_w \gg 1$:
$I(t_w,z) \simeq A t_w^{\alpha}  z^{\beta} \left[1-\frac{B}{(zt_w)^{\gamma}} \right]$.
Then we need to consider two separate cases for the evaluation of 
${\hat{\cal P}}^{(k,k)} (\tau, t_w, z)$ from relation (\ref{laplace_k}).

\subsubsection{{\bf (a1) Case \ $0 < k < \frac{1+\mu}{2(1-\mu)}$}}

For $0 < k < \frac{1+\mu}{2(1-\mu)}$ (i.e $\beta <0$), one gets for 
$zt_w \gg 1$:
${\hat{\cal P}}^{(k,k)} (\tau, t_w, z) \simeq 
A \frac{C_N^k}{ 2 \pi \tau}
\left[ \frac{c(\mu) \tau}{\tau_0^{\mu}}\right]^k
t_w^{\alpha}  z^{\beta}$.

Therefore, 
${\cal P}^{(k,k)} (\tau, t_w, T)$ is a power-law:

$$
{\cal P}^{(k,k)} (\tau, t_w, T) \simeq 
\frac{C_N^k}{ 2 \pi \tau}
\left[ \frac{c(\mu) \tau}{\tau_0^{\mu}}\right]^k
\frac{A t_w^{\alpha}}{\Gamma(-\beta)} \frac{1}{T^{1+\beta}} 
$$

\subsubsection{{\bf (a2) Case \ $ \frac{1+\mu}{2(1-\mu)}< k < \frac{1}{1-\mu}$}}

In this case, we find that
${\hat{\cal P}}^{(k,k)} (\tau, t_w, z)  \simeq
\exp \left[ -\frac{{\cal K}(t_w,\tau)}{(zt_w)^{\beta}}\right]$,
with
${\cal K}(t_w,\tau)=\frac{2 \pi \tau}{C_N^k}
\left[ \frac{\tau_0^{\mu}}{c(\mu) \tau}\right]^k
\frac{1}{t_w^{\alpha-\beta}}$.

The evaluation of the inverse 
Laplace transform leads to:

$$
{\cal P}^{(k,k)} (\tau, t_w, T) \simeq  
e^{\cos (\frac{\pi}{1+\beta} )
(\beta {\cal K}(t_w,\tau))^{\frac{1}{1+\beta}} 
\left(\frac{T}{t_w}\right)^{\frac{\beta}{1+\beta}}
}
$$

which is a stretched exponential since $\cos ( \frac{\pi}{1+\beta}) <0$.

\subsubsection{{\bf (b) Case \ $k(1-\mu) >1$}}

Finally, we consider the case $k(1-\mu) >1$: then we have to introduce the lower-cut-off
$\tau_0$ in   $I(t_w,z)$, which leads to:

$$I(t_w,z) \simeq z^{(1-\mu)k-1}
\int_{\tau_0 z}^{\infty} \frac{du}{u^{(1-\mu)k}} e^{-u} 
e^{K(k) \left( \frac{u}{zt_w} \right)^{\frac{1-\mu}{2}}}.$$

The saddle point solution is 
$u^*=[(1-\mu) K(k)/2]^{\frac{2}{1+\mu}} (zt_w)^{\frac{\mu -1}{1+\mu}}$. In the limit 
of very large $zt_w$, one will at some point reach the situation where $u^* < \tau_0 z$.
Hence,
for $zt_w \gg 1$, the integral will be best evaluated by its lower cut-off value:
$I(t_w,z) \simeq \frac{1}{z} \frac{1}{\tau_0^{(1-\mu)k}}
e^{-\tau_0 z} e^{K(k) \left(\frac{\tau_0}{t_w}\right)^{\frac{1-\mu}{2}}}$.
Since $T \gg \tau_0$, we finally get:
${\hat{\cal P}}^{(k,k)} (\tau, t_w, z)  \simeq
\frac{{\cal C}(t_w,\tau)}{z}$, where
${\cal C}(t_w,\tau)=\frac{C_N^k}{ 2 \pi \tau}
\left[ \frac{c(\mu) \tau}{\tau_0^{\mu}}\right]^k
\frac{e^{K(k)\left(\frac{\tau_0}{t_w}\right)^{\frac{1-\mu}{2}}}}{\tau_0^{(1-\mu)k}}$.

Finally,
${\cal P}^{(k,k)} (\tau, t_w, T)$ is simply an exponential, with a weak dependence on $t_w$
(which vanishes for large $t_w$):

$$
{\cal P}^{(k,k)} (\tau, t_w, T) \simeq 
{\cal C}(t_w,\tau)  e^{-{\cal C}(t_w,\tau) T} 
$$

\section {APPENDIX B}

For the computation of ${\cal P}^{(k',k'')}_0 (\tau, t_w, T)$,
we will proceed in the same way as in the last section for 
${\cal P}^{(k,k)}_0 (\tau, t_w, T)$.
Using again combinatorial arguments, we can write that:

\begin{eqnarray*}
 & & {\cal P}_0^{(k',k'')} (\tau, t_w, T)= 
 \frac{C_N^{k'}}{\tau} \sum_{n=n_{inf}}^{n_{sup}}
C_{k'}^n C_{N-k'}^{k''-k'+n} \left[ \tau S(T) \right]^{k'-n} \\
& & \left[ 1-\tau S(T) \right]^{n}
\left[ \tau S(t_w+T) \right]^{k''-k'+n}
\left[ 1-\tau S(t_w+T) \right]^{N-k''-n}\\
\end{eqnarray*}

\noindent
where $n_{inf}={\rm sup}(0,k'-k'')$ and $n_{sup}={\rm inf}(k', N-k'')$.

As before, we keep only the leading order in $\tau$ and go to the large $N$ limit by
introducing $x'=\frac{k'}{N}$, $x''=\frac{k''}{N}$ and $y=\frac{n}{N}$. This leads to:

$$ {\cal P}_0^{(k',k'')} (\tau, t_w, T)\simeq 
\frac{\tau^{k''-1}}{(2\pi)^{3/2} N^{1/2}}
\int_{x_{inf}}^{x_{sup}} dy \ 
e^{-N {\cal S}(x',x'',y)},$$

where

\begin{eqnarray*}
& &{\cal S}(x',x'',y)=y \ln y 
+(x'' -x' +y) \ln (x'' -x' +y)\\
&+& (x'-y) \ln (x'-y)
+(1- x'' -y) \ln (1- x'' -y) \\
&-& (x'-y) \ln S(T)
-(x'' -x' +y) \ln S(t_w+T).
\end{eqnarray*}

We use a saddle-point approximation to minimize ${\cal S}(x',x'',y)$:
$\frac{\partial {\cal S}}{\partial y} |_{y^*} =0$ leads to:

$$(x'-y^*)(1-x''-y^*)=y^* (x''-x'+y) \gamma(t_w,T)$$

which is the generalization of equation (\ref{eq_y*}).

The positive solution is therefore:

\begin{eqnarray*}
& & y^*=\frac{1}{2(\gamma-1)} [ -1-(\gamma -1) (x'' -x')\nonumber\\
& & + \sqrt{[1+(\gamma-1)(x''-x')]^2 +4x'(1-x'')(\gamma-1)}] 
\end{eqnarray*}

\noindent
and it is easy to check that $x_{inf} \leq y^* \leq x_{sup}$.
Note that in this equation and in the following, we drop the $(t_w,T)$ dependence
of $\gamma(t_w,T)$ for convenience.


We now come back to the calculation of
${\cal P}_0^{(k)} (\tau, t_w, T)$  through equation (\ref{sum}). For large $N$, we have:

$${\cal P}_0^{(k)} (\tau, t_w, T) \simeq N^2
\int_x^1dx' \int_x^1 dx''
{\cal P}_0^{(k',k'')} (\tau, t_w, T).$$

Direct integration is not possible here, so we use again a saddle point treatment of the double
integral. The saddle points $x'^*$ and $x''^*$ are solutions of the set of equations formed
by:
$\frac{\partial {\cal S}}{\partial x'} |_{(x'^*,x''^*)} =0$ and
$\frac{\partial {\cal S}}{\partial x''} |_{(x'^*, x''^*)} =0$, or:

\begin{equation}
\frac{\partial y^*}{\partial x'}
\ln \frac{y^*(x''-x'+y^*) \gamma}{(1-x''-y^*)(x'-y^*)}= 
\ln \frac{\gamma (x''-x'+y^*)}{(x'-y^*)} \label{dif1}
\end{equation}

\begin{equation}
\frac{\partial y^*}{\partial x''}
\ln \frac{y^*(x''-x'+y^*) \gamma}{(1-x''-y^*)(x'-y^*)}=
\ln \frac{\tau S(T+t_w)(1-x''-y^*)}{x''-x'+y^*} \label{dif2}
\end{equation}

Here, $\frac{\partial y^*}{\partial x'}|_{(x'^*,x''^*)}$ and
$\frac{\partial y^*}{\partial x''}|_{(x'^*,x''^*)}$ can be calculated using 
the expression for $y^*$.

At this point, taking the $\frac{T}{t_w} \ll 1$ limit enables us to make simplifications in
equations (\ref{dif1}) and (\ref{dif2}). A rather lengthy calculation leads to the conclusion that there is no
couple of solutions $(x'^*, x''^*)$ compatible with equations 
(\ref{dif1}) and (\ref{dif2}), and such that $x'^*
\neq x''^*$.

Consequently, we keep only the terms such that $k'=k''$ in the sum of (\ref{sum}), and try to find
the value
for $k'$ that maximizes this sum, knowing the expression of 
${\cal P}_0^{(k',k')} (\tau, t_w, T)$
from  section  
IV.3:

\begin{eqnarray*}
& &{\cal P}_0^{(k)} (\tau, t_w, T) \propto \int_x^1 dx'  \ 
\exp \{ N [-x' \ln x'   \nonumber \\
&-&(1-x') \ln (1-x') + x' \ln S(T) +2 (\frac{x'(1-x')}{\gamma(t_w,T)})^{1/2} ] \} 
\end{eqnarray*}

The saddle-point equation for $x'^*$ is now:

$$ \ln \left(\frac{x'^*}{1-x'^*} \right) =
\frac{1-2x'^*}{\sqrt{x'^*(1-x'^*) \gamma(t_w,T)}} + \ln [\tau S(T)].$$

If $\tau S(T)=1$, this equation has the only solution $x'^*=\frac{1}{2}$. For $\tau S(T) > 1$,
there is a solution $x'^* > \frac{1}{2}$, and for 
$\tau S(T) < 1$,
there is a solution $x'^* < \frac{1}{2}$. In general, $x^*= {\rm sup} (x,x'^*)$ 
will maximize
the expression for ${\cal P}_0^{(k)} (\tau, t_w, T)$, 
and can be estimated numerically. In particular, if $\tau S(T)< 1$ and $k> N/2$, we
have $x^*=x$.

Finally,

$${\cal P}_0^{(k)} (\tau, t_w, T)={\cal P}_0^{(k^*)} (\tau, t_w, T),$$

$${\cal P}^{(k)} (\tau, t_w, T)={\cal P}^{(k^*)} (\tau, t_w, T),$$

\noindent
where $k^*={\rm sup} (k,k'^*)$.

\bibliographystyle{unsrt}

\end{document}